\documentclass[%
 reprint, 
 amsmath,amssymb,
 aps, 
]{revtex4-2}

\usepackage{graphicx}
\graphicspath{ {./images/} }
\usepackage{dcolumn}
\usepackage{bm}
\usepackage{pgf}
\usepackage{pgfplots} 
\usepackage{braket}
\usepackage{siunitx}
\usepackage{xcolor}
\usepackage[tikz]{bclogo}
\usepackage{graphicx,array,multirow}
\usepackage{svg,float}
\usepackage{tabularx}
\usepackage{hyperref}
\usetikzlibrary{arrows.meta, positioning, decorations.pathreplacing}
\usepackage{soul}
\usepackage{gensymb}
\usepackage{makecell}

\begin{document}

\preprint{APS/123-QED}

\newcommand{\vkeff}{\vec{k}_{\text{eff}}}
\newcommand{\keff}{k_{\text{eff}}}
\newcommand{\red}[1]{\textcolor{red}{#1}}
\newcommand{\blue}[1]{\textcolor{blue}{#1}}

\title{\textbf{Stroboscopic Raman Spectroscopy of Atom Optics in Quasi-Bragg Regime} 
}%

\author{Joel Gomes Baptista}
\altaffiliation[Presently at ]{Leibniz Universität Hannover, Institut für Quantenoptik, Welfengarten 1, 30167 Hannover, Germany}
\author{Louis Pagot}%
\author{Sébastien Merlet}%
\author{Leonid A. Sidorenkov}%
\author{Franck Pereira dos Santos}%

\email{Contact author: franck.pereira@obspm.fr}
\affiliation{%
  LTE, Observatoire de Paris - Université PSL, Sorbonne Université, Université de Lille, LNE, CNRS, Paris, France
}%

\date{\today}

\begin{abstract}
Quasi-Bragg regime is a good compromise for large-momentum-transfer atom interferometry, allowing for scaling up the interferometric area, while constraining the population of unwanted states. Separation of momentum states via standard time of flight methods, however, can be challenging when using laser-cooled atoms, rather than ultracold atoms with sub-recoil velocity distribution. To overcome this limit, we use Raman spectroscopy for stroboscopic sampling of the atomic state evolution in momentum space during the interrogating laser pulses. We quantitatively characterize atom optics employing two-photon ($2\hbar k$) and multi-photon ($6\hbar k$) Bragg transitions, the latter being optionally enhanced with optimal control protocol. We closely match the observed dynamics of the atomic state with simulations. Finally, we perform momentum spectroscopy of the output states in a $6\hbar k$ Bragg gravimeter. 
\end{abstract}

\maketitle

 \section{\label{sec:Introduction} Introduction\protect\\}
 Bragg diffraction is a central tool in atom interferometry, in particular for large momentum transfer (LMT) interferometer where increasing the separation between the arms enhances the sensitivity \cite{chiow$102ensuremathhbark$LargeArea2011,kovachyQuantumSuperpositionHalfmetre2015,mazzoniLargemomentumtransferBraggInterferometer2015,damicoBraggInterferometerGravity2016,geigerHighaccuracyInertialMeasurements2020,zhangUltrahighSensitivityBraggAtom2023}. In practice, operating in the quasi-Bragg regime leads to population transfer into parasitic momentum states, so that the quantum state dynamics can no longer be described as a two-level atomic system \cite{buchner2003,muellerAtomwaveDiffractionRamanNath2008,beguinCharacterizationAtomInterferometer2022}.
 The effect of those parasitic states is especially relevant for atomic interferometers based on laser-cooled atoms, since finite velocity spread and coupling inhomogeneities induced by ballistic expansion reduce the fidelity of high-order Bragg pulses.

A major experimental drawback of using laser cooled, rather than evaporatively cooled ultracold atom sources, lies in the difficulty of reaching sufficient spatial separation via time of flight (TOF) to resolve adjacent Bragg states, which prevents a direct characterization of the Bragg dynamics during the diffraction process or throughout the interferometric sequence. To mitigate this effect, one can increase the separation between momentum states using a Bloch elevator lattice \cite{altinPrecisionAtomicGravimeter2013,picconSeparatingOutputPorts2022}. However, the method suffers from imperfect transport efficiency, and is better suited for quasi-ideal two-states systems. To address these limitations, we implement a detection scheme based on velocity-selective Raman spectroscopy, similar to the one reported in Refs.~\cite{Cheng2018mrd,Beaufils2022cas} which enables to monitor the atomic momentum distribution at any given moment during the pulse. This approach allows us to reconstruct the evolution of the atomic state, compare the measured dynamics with numerical simulations, and quantify the impact of the coupling dispersion both during individual diffraction pulses and at the output of a complete interferometer.

 In this work, we apply stroboscopic Raman spectroscopy for characterizing first order ($2\hbar k$) and third order ($6\hbar k$) Bragg transitions in a cold-atom gradiometer. We show that Raman spectroscopy detection provides an efficient probe of the complex Bragg dynamics when optimal control transfer (OCT) techniques \cite{khanejaOptimalControlCoupled2005,Boscain2021, saywellOptimalControlRaman2020} are used to optimize the laser phase temporal evolution during the diffraction process. 
 Finally, we extend the analysis to a complete Mach-Zehnder interferometer sequence, where state-resolved detection reveals interference fringes for each Bragg state and allows for quantifying the contribution of parasitic states to the degradation of interferometer performance as the diffraction order increases.

\section{\label{sec:Gradiometer} Experimental setup\protect\\}

Our apparatus, previously described in detail in \cite{caldaniSimultaneousAccurateDetermination2019}, is a dual gravity sensor measuring both the gravity acceleration and its gradient. The sensor head comprises a drop chamber with two (top and bottom) vacuum chambers for the production of the atomic sources, which are connected to a common vertical tube and separated by a distance of \qty{1}{\meter}.
Each source chamber hosts an in-vacuum mirror for the simultaneous realization of two surface 3D-MOTs. 
A collimated laser beam of a waist $w=\qty{4}{\mm}$ enters at the bottom of the drop chamber and propagates upward, thus crossing both source chambers. It is then retroreflected onto a mirror located outside the vacuum chamber, forming an optical lattice. Such configuration enables simultaneous interrogation of both atomic clouds with common light pulses. The interrogating light is brought to the chamber via an optical fiber from the output of a laser system, described in detail in \cite{sarkarSimpleRobustArchitecture2022}. This versatile laser system allows for realizing different atom-optics operations during the measurement cycle, including: velocity selection and spectroscopy with Raman pulses, optional atom launching and wavepacket separation with Bloch elevators and interferometry with Bragg pulses.  

The experimental sequence starts with the loading of the 3D-MOTs from the intense flux of two independent 2D-MOTs for a duration of \qty{670}{\milli\second}. About $10^{8}$ atoms of $^{87}$Rb are then laser-cooled to approximately \qty{2}{\micro\kelvin} via far detuned optical molasses and prepared in the $F=2$ hyperfine ground state before being released in free fall. 
After a few ms of TOF, we apply a square \qty{70}{\micro\second} long Raman pulse, out-coupling the atoms having low vertical velocity spread into $\ket{F=1,m_F=0}$ hyperfine ground state that will further enter interferometric sequence. This process reduces the vertical rms velocity spread from $\sigma_v=2\,v_r$ to about $\sigma_v=0.3\,v_r$, $v_r$ being atom recoil velocity, while the transverse velocity distribution remains unchanged. 

To simultaneously probe gravitational accelerations affecting both atom clouds, we use a symmetric Mach-Zehnder sequence of three (beamsplitter-mirror-beamsplitter) Bragg pulses separated by a time interval $T$. We use Gaussian pulses of rms durations $\sigma_{t} = \qty{7}{\micro\second}$ and $\qty{14}{\micro\second}$ for the beamsplitter and mirror pulses, respectively. 
We resonantly drive either $2\hbar k$ or $6\hbar k$ Bragg transitions by adjusting the two-photon detuning, and maximize the efficiency of the pulses using optimal Rabi couplings of $\Omega_R/2\pi$=14.25~kHz, for $2\hbar k$ (diffraction order $n=1$) and 57~kHz, for $6 \hbar k$ ($n=3$). This places our experiment in the quasi-Bragg regime, where excitation of parasitic states is negligible in the case of resonant adiabatic pulses.


However, inhomogeneities in the detuning, due to the Doppler effect, and in the coupling amplitude, due to the finite sizes of the laser beam and of the atomic clouds, 
reduce the fidelity of atom-optics and increase the population fraction of parasitic momentum states participating in the interferometric process. To minimize the impact of transverse time-of-flight expansion of atomic clouds, we restrict in this study the interferometric cycle constant to $T=\qty{1}{\milli\second}$. Experimentally, we find maximum efficiencies for the mirror pulses of $\qty{60}{\percent}$ ($n=1$) and $\qty{40}{\percent}$ ($n=3$). The respective interferometric contrast is preserved at the levels of $\approx \qty{40}{\percent}$ and $\approx\qty{20}{\percent}$. 


The atoms are finally detected at the very bottom of the drop chamber, where the populations in the two hyperfine ground levels are measured via resonant fluorescence using three successive retroreflected light sheets. The detection sheets are shaped by rectangular slits with a common width of \qty{15}{\milli\meter}, and heights of \qty{5}{\milli\meter} for the first and the third, and \qty{1}{\milli\meter} for the second.
The first light sheet detects the atoms in $\ket{5S_{1/2},\, F = 2}$ ground state as they are driven on the cycling transition $\ket{5S_{1/2},\, F = 2} \rightarrow \ket{5P_{3/2},\, F' = 3}$. At the bottom of this light sheet, the retroreflected beam is shadowed, in order to push the detected atoms away by radiation pressure. 
Atoms in the state $\ket{5S_{1/2},\, F = 1}$ keep falling and get transferred to the $\ket{5S_{1/2},\, F = 2}$ state in the second repumping light sheet. Finally, they are detected by the last light sheet resonant with the cooling transition.

\section{\label{sec:Momentum-resolved detection} Momentum-resolved detection\protect\\}
\begin{figure}[tbp!]
        \centering
        \includegraphics[width=\linewidth]{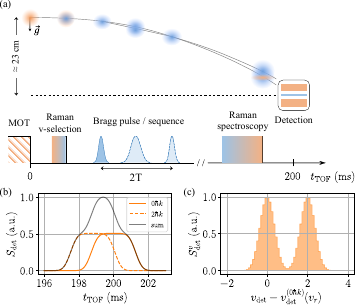}
    \caption{(a) Scheme of the experimental sequence used in this work and simulated signals for the detection using 5 mm light sheet of balanced-population in 0 and 2 $\hbar k$ states using (b) time-of-flight and (c) Raman spectroscopy with total $0.2\,v_r$-resolution (columns, see text). The orange (blue) color encodes atomic population in $\ket{F=2}$ ($\ket{F=1}$) state. The 2nd and the 3rd Bragg pulses (light blue-filled with dashed envelope) are optionally used to study a complete closed interferometer.}
    \label{fig:Figure1}
\end{figure}
In this work, we explore the methods for momentum-resolved detection of Bragg states, pertinent to a compact gravimeter configuration (see Figure~\ref{fig:Figure1}a), by employing only the bottom atomic source with a free-fall distance of about \qty{23}{\centi\meter}, given by the location of its 3D-MOT above the detection region. At a corresponding TOF limited to 200 ms, the spatial separation of adjacent momentum states differing by a velocity of $2\,v_r\approx 12~$mm/s is approximately $\qty{2.4}{\milli\meter}$. Given a typical rms-width of $\qty{0.6}{\milli\meter}$ of the detected atomic cloud, dominated by its initial size upon release from the optical molasses \cite{Sidorenkov2020tma}, one could in principle resolve these states using sub-millimeter light sheets, at the price of drastic reduction of the fluorescence signal. 
Conversely, an order of magnitude longer free-fall distance would be required to spatially resolve the adjacent momentum states in our present 5 mm light sheet detection system.  



To sufficiently separate \textit{two} momentum states, we have previously used a Bloch separator method \cite{picconSeparatingOutputPorts2022}, accelerating one of the components with a resonant optical lattice 
before detection. The finite efficiency of the separator, however, leads to a residual mixing between the two populations, which complicates the analysis of the Bragg dynamics and reduces the contrast of the interferometer. This method is therefore hardly applicable for LMT Bragg interferometry as it involves complex dynamics of \textit{multiple} momentum states $\{n\hbar k$\}. 

An alternative method introduced in \cite{Cheng2018mrd} and adapted here consists in using an additional Raman pulse transferring the population of a given momentum state $n\hbar k$ (or a narrow velocity sub-group in its vicinity) to another hyperfine state and its subsequent detection with the resonant light sheet. In this case, one benefits from the full-amplitude fluorescence signal, while the controlled duration and frequency detuning of the Raman pulse define the momentum resolution independently of the size of the detection light sheet and total time of flight. 

As an illustration, we show in Figure~\ref{fig:Figure1} a simulated detection of balanced-superposition state of 0 and 2 $\hbar k$ with conventional time-of-flight (panel b) and corresponding velocity spectrum sampled with Raman pulse selecting rms of $0.1\,v_r$ (panel c), three times longer than the one used in preparation phase - modeled here for simplicity with a flat window of $0.2\,v_r$-width. The insufficient separation between two momentum states in TOF distorts the detected signal prevents a correct estimation of corresponding populations. On the contrary, the spectrum recorded using an additional Raman pulse allows to even visualize details of the momentum distributions for both states.

\section{\label{sec:Raman_spectroscopy} In-pulse evolution of Bragg states\protect\\}

\subsection{Stroboscopic Raman spectroscopy}

Using the additional Raman pulse, we can scan the velocity distribution across the Bragg states of momenta $\ket{p_0+2n\hbar k}$. To follow the in-pulse dynamics of the atomic state, we abruptly terminate the Bragg pulse by switching off the RF signal driving the AOM that controls the coupling amplitude. The effective truncation time is limited by the AOM fall time of a few \qty{100}{\nano\second}, much shorter than the characteristic evolution time of the system ($\tau_{\mathrm{c}}=1/\Omega_R\approx \qty{10}{\micro\second}$). After cutting the Bragg pulse, we shine a Raman pulse, that is three times longer than the velocity-selection pulse, transferring a selected resonant $\approx2\times 0.1\,v_r$-wide velocity class to the $\ket{F=2}$ state, subsequently detected by the first light sheet. We then repeat the experiment while changing the Raman detuning and reconstruct the full spectrum of the "frozen" evolution at a given cutting moment. 

\begin{figure}[tbp!]
        \centering
        \includegraphics[width=\linewidth]{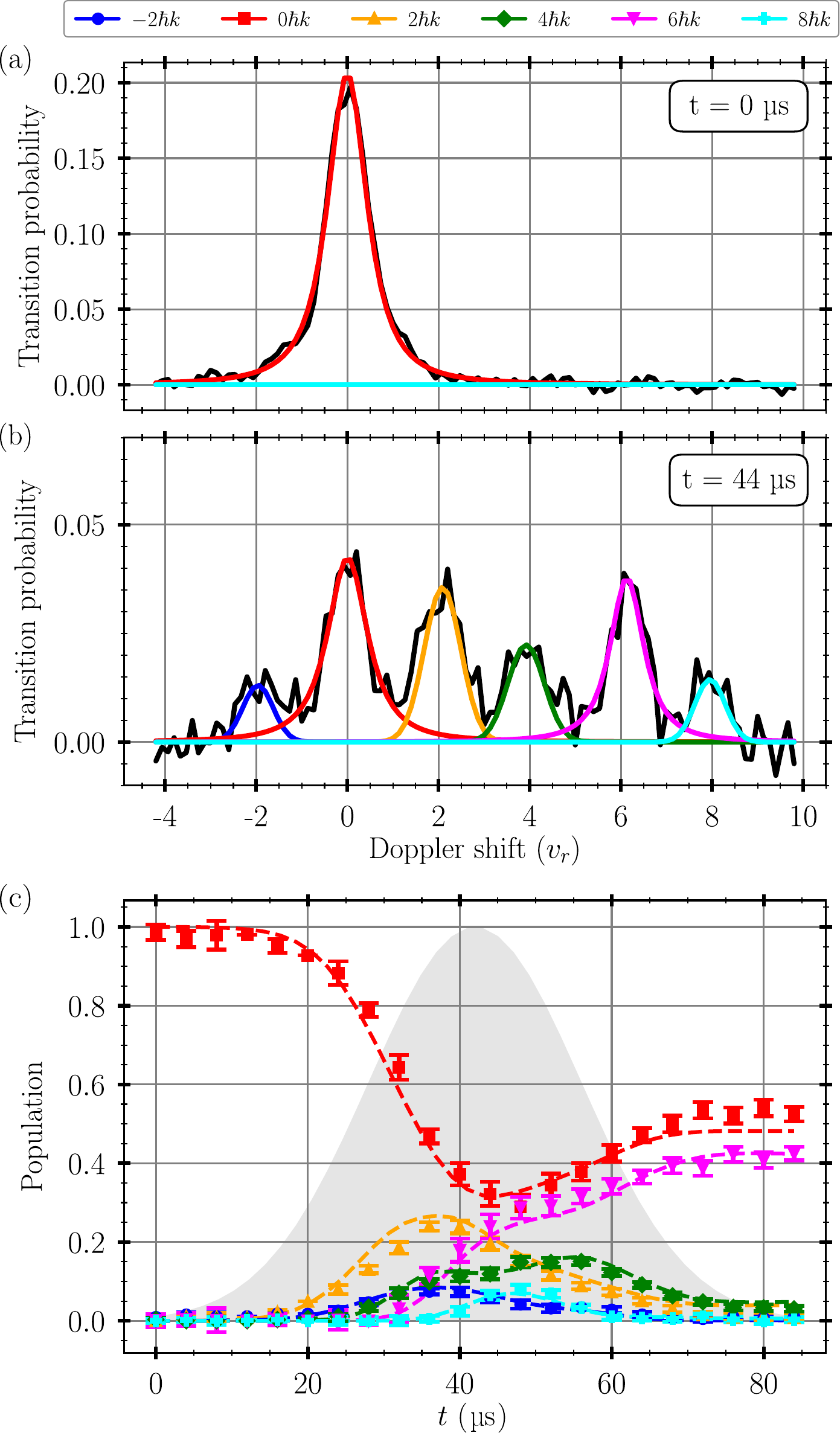}
    \caption{Raman spectra of the momentum states (a) just before ($t=\qty{0}{\micro\second}$) and (b) in the middle ($\qty{40}{\micro\second}$) of $6\hbar k$  Bragg mirror interrogation pulse. 
    The offsets due to spontaneous emission and experimental imperfections have been fitted and removed from the shown spectra, for clarity.
    (c) Complete evolution of the populations in significant Bragg states during $6\hbar k$  Bragg mirror pulse. Symbols: the populations computed as the normalized fitted areas of the peaks in the measured spectra exampled in panels a and b. Dashed lines: numerical simulation (see text). Gray shaded area marks the Gaussian envelope of the coupling amplitude.}
    \label{fig:Figure2}
\end{figure}


Figure~\ref{fig:Figure2} displays typical Raman spectra obtained at the start (panel a) and in the middle (panel b) of the $6\hbar k$ Bragg mirror pulse. The peaks in the spectra are labeled according to their mean momentum difference with respect to the initial state $\ket{p_0}$ ($0\hbar k$). Panel (a) shows the measured velocity distribution of the initial state which reveals significant characteristic Lorentzian-like wings. Here we expect observing a sinc-function spectrum originating from the square velocity-selectivity pulse. However, its spectral lobes appear smoothened, due to the convolution with the three-times thinner sinc-function of the spectroscopy pulse. For this reason, we use an empiric q-Gaussian (also called Lorentz-B, for instance in ~\cite{Altorio2020ata}) function for fitting the peaks of initial and target ($6\hbar k$) states, while the other peaks are fitted with Gaussian functions (see panel b). By integrating the fitted area under each peak, we extract the corresponding populations thus evaluating partial contributions to the diffraction process.

We then perform such scans of the velocity distribution every \qty{4}{\micro\second} and reconstruct in detail the evolution of the Bragg states during the mirror pulse, which is presented in the panel (c) of Figure \ref{fig:Figure2}. We observe significant population in all momentum states, varying rather smoothly during the Bragg diffraction pulse. At the end of the pulse, all but initial and target states almost vanish, thus underlining the benefits of quasi-Bragg regime. However, we may note a stagnation of the main states' populations at the values around 50\% without crossing each other, i. e. without an inversion expected for a $\pi$-pulse (mirror). We attribute this feature to the strong impact of coupling inhomogeneities in diffraction process, which motivates us to confront the measurements with corresponding numerical simulations, before proposing mitigation protocols. 

\subsection{Numerical model}
The numerical simulation of the Bragg dynamics is based on the resolution of the Schrödinger equation of an atom in a Bragg optical lattice under gravity~\cite{beguinCharacterizationAtomInterferometer2022} (see Appendix B). 
To realistically reproduce the experimental conditions, the simulation averages over two sources of inhomogeneity: the initial velocity distribution, given by the spectral width of the Raman velocity selection pulse, and the spatial distribution of the two-photon Rabi frequency $\Omega_R$.



To account for the latter contribution, we consider an atomic cloud with initial (at the moment of first Bragg pulse) Gaussian distribution of standard deviation $\sigma_r$, illuminated by a Gaussian beam of  waist ($1/e^2$-radius) $w$. The two-photon Rabi frequency then follows the spatial intensity profile of the beam $\Omega_R(\rho) = \Omega_R^0 \exp\!\left(-2\rho^2/w^2\right)$, where $\rho$ is the transverse distance to the beam axis and $\Omega_R^{0}$ denotes the peak Rabi frequency. Averaging this quantity over the spatial distribution of the atomic cloud leads to a coupling distribution characterized by a mean value $\langle \Omega_R \rangle$ and a rms width $\sigma_{\Omega_R}$. We define the effective relative coupling inhomogeneity as
\begin{equation}
\chi = \frac{\sigma_{\Omega_R}}{\langle \Omega_R \rangle}
= \sqrt{\frac{\langle \Omega_R^2\rangle}{\langle \Omega_R\rangle^2}-1}= \frac{ r_C^2}{\sqrt{1 + 2 r_C^2}}
\end{equation}
where $r_C = 2\sigma_r / w$ is the ratio between the transverse rms diameter of the atomic cloud and the beam waist. For a cloud of rms size $\sigma_r=\qty{0.5}{\milli\meter}$, $r_C = \qty{0.25}{}$, which corresponds to an effective relative coupling inhomogeneity $\chi=\qty{5.9}{\percent}$.

To match the experimental Raman spectroscopy measurements, the simulation parameters $\Omega^0_R$ and $\sigma_r$ are treated as fitting parameters. The details of the fitting procedure can be found in Appendix B. Results of the simulation with adjusted parameters are displayed in Figure \ref{fig:Figure2}c as dashed lines. A good agreement with experimental data in Figure~\ref{fig:Figure2} is found for the following fit parameters $\Omega^0_R/2\pi=66\pm2$~kHz and $\sigma_r=1.23\pm0.06$~mm. This corresponds to a strong effective relative coupling inhomogeneity of $\chi=29\pm 2~\%$, that may be explained by several factors. First, the actual shape of the cloud upon release from optical molasses could be an ellipsoid stretched along the weaker-gradient axis of the MOT at $45\degree$ with respect to vertical, thus giving a larger initial horizontal spread in one of the axes. Second, the simulation doesn't account for diffraction of the interrogating beam on the edges of surface-trap mirrors that would create a stronger transverse variation than a simple Gaussian profile. The corresponding excessive inhomogeneity could thus be absorbed by effectively increased fitted cloud size.

The results of stroboscopic spectroscopy, matched to the simulation of atomic state evolution, allow for an accurate characterization of atomic source dispersion responsible for coupling inhomogeneity that strongly limits the efficiency of LMT atom-optics. We demonstrate below effective measures to circumvent this limitations being explored with the present spectroscopy and the simulation toolbox.


\subsection{\label{sec:Bragg_analysis} Quantum control of optimized mirror pulse}

To improve the efficiency of the Bragg process, we use optimal control theory to design atom-optics resilient to velocity and coupling inhomogeneities. We preserve the peak amplitude and the smooth temporal Gaussian envelope of the Rabi coupling pulse, but typically extend its duration three to fivefold. The only optimally-controlled parameter is the differential phase between the two laser beams forming the Bragg optical lattice.
In the case of a two-level system, this phase would correspond to an equatorial angle of the torque field vector in the Bloch sphere picture.

\begin{figure}[tbh!]
    \centering
    \includegraphics[width=\linewidth]    {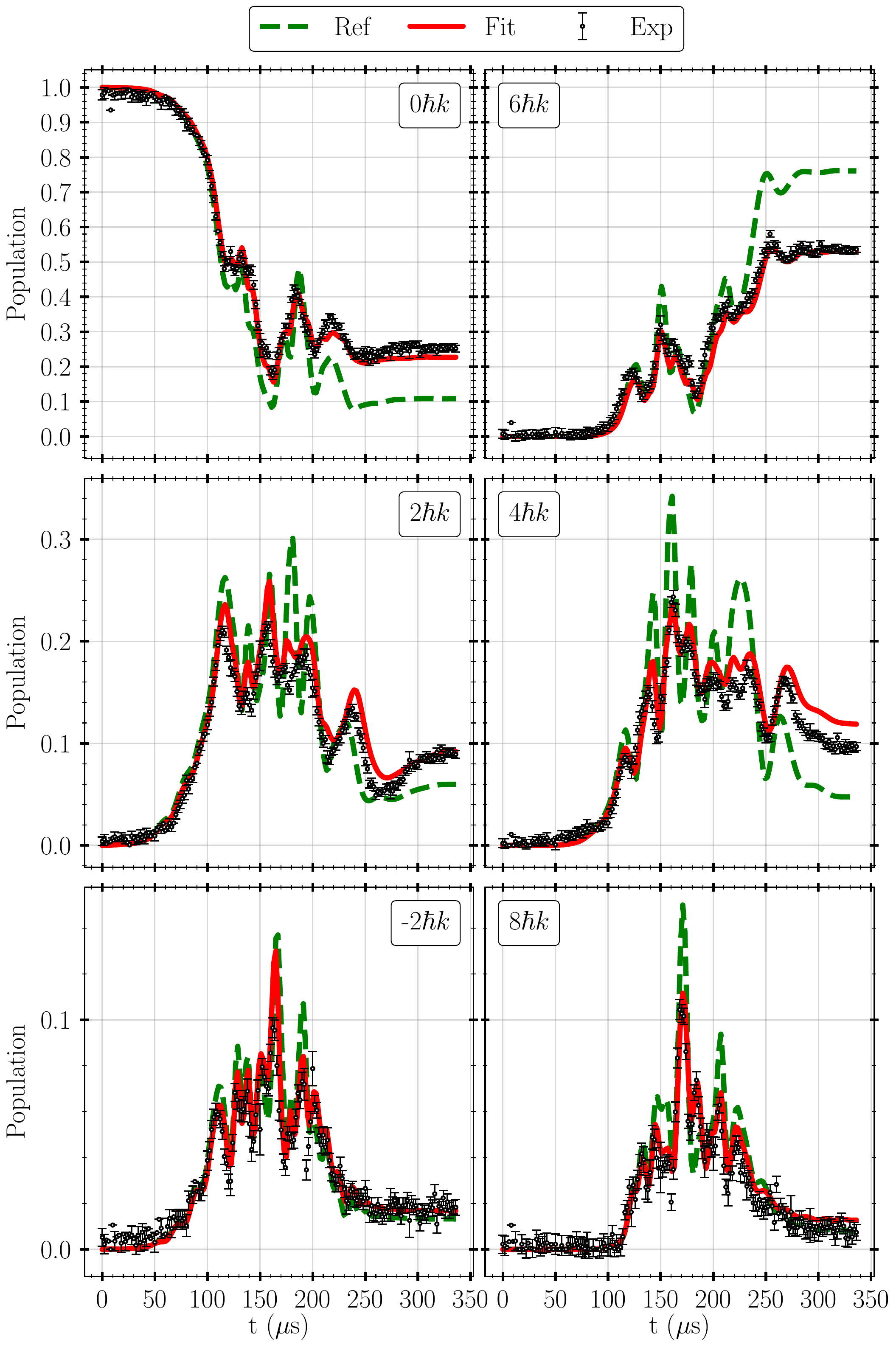}
    \caption{Evolution of the momentum states in the $6\hbar k$ Bragg mirror pulse optimized with OCT, recorded using stroboscopic Raman spectroscopy (symbols). The data is confronted with simulation for reference (dashed green line) and fitted (solid red line) parameter configurations, see text.}
    \label{fig:evolution_spectro_pi_pulse_6hk_OCT}
\end{figure}

The black symbols in Figure~\ref{fig:evolution_spectro_pi_pulse_6hk_OCT} display the evolution of the six most populated momentum states for such an optimized Bragg mirror pulse. Here, we resolve fine local variations in states' populations, which demonstrate an overall more complex dynamic than in the case of the conventional $6\hbar k$ pulse shown in Figure~\ref{fig:Figure2}c. The OCT allows to significantly lower the population in initial ($0\hbar k$) state and increase that of the target one ($6\hbar k$), manifesting the inversion at the end of the pulse. However, about 20\% of the population remains in unwanted momentum states. 

We account for an OCT phase pattern in the previously described simulation approach and obtain numerical results corresponding to the reference case of an atomic cloud with an rms size $\sigma_r=\qty{0.5}{\milli\meter}$ and $\Omega^0_R/2\pi=57$~kHz (dashed green lines) and to the case of fitted values of $\sigma_r$ and $\Omega_R^0$ (solid red lines). Adjusting the parameters allows to accurately reproduce the in-pulse dynamics associated to the quantum control protocols, which furthermore enhances our confidence in the simulation. 

\subsection{\label{sec:Bragg_analysis} Discussion}
Using identical preparation, detection, analysis and numerical simulation routine, we have studied various Bragg diffraction pulses including $n=3$ mirror (conventional and with OCT), as well as $n=3$ beamsplitter with OCT and $n=1$ conventional mirror pulse. The latter two data sets are presented in Figures~\ref{fig:evolution_spectro_pi_over_2_pulse_6hk_OCT} and \ref{fig:evolution_spectro_pi_pulse_2hk_no_OCT} of Appendix. The relative inhomogeneities obtained from adjusting the values of $\Omega_R$ and $\sigma_r$ to best match each experimental data are reported in Table~\ref{tab:optimized_value}.


For both first and third order Bragg transitions, the agreement between simulated and measured population dynamics requires increasing the effective cloud size and the peak Rabi frequency with respect to their nominal values, which leads to a significant increase of the relative coupling inhomogeneity (mean value calculated from the Table~\ref{tab:optimized_value}: $\chi=\qty{33}{\percent}$) compared to the value expected for the nominal parameters. This larger dispersion indicates that the atoms experience larger intensity inhomogeneities than expected from a pure Gaussian beam.
\begin{table}[t!]
\caption{Fitted parameters and corresponding relative inhomogeneities obtained from the adjustment of the atomic population dynamics for various Bragg pulses (M = Mirror,  Bs = Beamsplitter).}
\label{tab:optimized_value}
\begin{tabular}{|c|c|c|c|c|}
\hline
\multirow{2}{*}{\makecell{ Diffraction \\ pulse type } } &{$n = 1$} & \multicolumn{3}{c|}{$n = 3$} \\ \cline{2-5}
 & {M} & {M} & {M, OCT} & {Bs, OCT} \\
\hline
 $\Omega_R^0/2\pi, \textrm{kHz}$ & $21.9(7)$ & $66(2)$ & $66(1)$ & $74.2(3)$ \\
\hline
 $\sigma_r, \textrm{mm}$ & $1.56(5)$ & $1.23(6)$ & $1.25(2)$ & $1.37(1)$ \\
\hline
 $\chi, \%$ & 40(2) & 29(2) & 29.3(7) & 33.7(4) \\
\hline
\end{tabular}
\end{table}

The good agreement between the experimental data and the optimized simulation confirms the validity of the method, which allows us to accurately capture the multimode dynamics of the quasi-Bragg regime for high diffraction orders. However, the match is not perfect and could be improved by different means. 

First, the Raman spectroscopy pulse is a square pulse only three times longer and weaker than the velocity selection pulse. A better momentum resolution could be achieved by increasing the duration of this pulse. In addition, using an adapted temporal shape (Gaussian, Gsinc or Blackman)~\cite{Fang2018itp} could suppress the spectral wings inherent to the sinc profile, providing a more clear spectroscopic signal. 

Second, the Bragg populations are deduced from fits to the measured spectrum. The choice of the fitting functions may be optimized. In particular, the use of the q-Gaussian profiles may fail to capture the dynamics of the spectral wings and lead to deviations from the real distribution. 

Last, the signal-to-noise ratio was limited in presented experiments, forcing to choose a relatively broad spectral probe that prevented the observation of the finest details of the momentum distribution. Improving the number of trapped and thus detected atoms, which was constrained by an aging tapered amplifier of our laser-cooling system, would increase the signal and allow for more selective spectroscopy pulses~\cite{Beaufils2022cas}.

We can furthermore envision applying the tested Raman detection protocol for a single-shot detection of the momentum states of interest in LMT Bragg diffraction, as opposed to acquiring the full spectrum, similarly to the demonstration in \cite{Cheng2018mrd} for $n=1$ and $n=2$ diffraction orders. For example, using two consecutive Raman pulses resonant with $\sigma_v=0.3\,v_r$ velocity classes centered on $0\hbar k$ and $6\hbar k$ and spectrally separated by sharply $6\,\nu_r$ ($\nu_r$ being recoil frequency of an atom), one can detect the populations of only main states using light-sheet fluorescence sufficiently separated in time-of-flight as all intermediate states remain transparent. The restrictions on the size of detection light sheet, the initial width of atomic cloud and the total time-of-flight become further relaxed in this case, as diffraction order of interest increases.

In this work, we have applied Raman spectroscopy for detecting the evolution of the complex momentum states optimized with OCT methods that control a single parameter, the laser phase. Being well matched to the simulation, this approach could be used to explore more complete OCT protocols that involve optimized variation of several parameters (such as phase, amplitude and frequency detuning of the driving field) - to study robustness ~\cite{saywellEnhancingSensitivityAtominterferometric2023, Louie2023rao, Baker2026rqc} and accuracy~\cite{Martinez2026dpf} of the LMT atom interferometers in quasi-Bragg regime.

\subsection{\label{sec:} Raman spectroscopy of a Bragg interferometer}
The Raman spectroscopy method can also be used to probe the output state of an atom interferometer sequence. We scan the output phase of the Mach-Zehnder gravimeter by varying the chirp $\alpha=d\nu/dt$ applied to the frequency difference of the Bragg lasers
around a mean value that compensates the Doppler detuning during free fall $\alpha_0=kg/\pi\approx -25.14$~MHz/s. For each chirp value, a Raman spectroscopy measurement is performed on the output ports of the interferometer, allowing us to resolve the population associated with each momentum state.

\begin{figure}[ht!]
    \centering
        \centering
        \includegraphics[width=\linewidth]{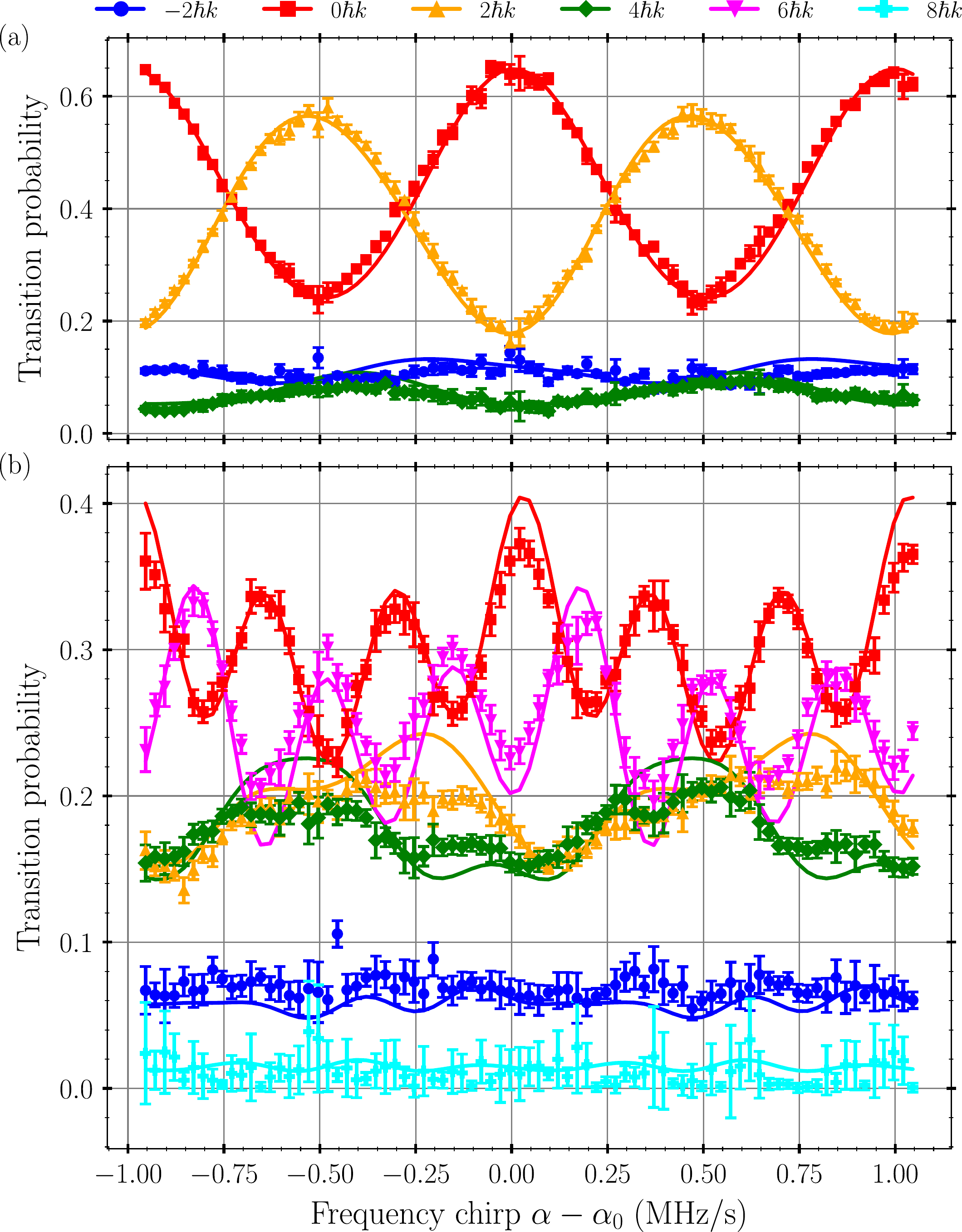}
    \caption{Fringe patterns for the interferometers employing (a) first and (b) third order Bragg atom-optics without OCT. Symbols are the populations of significant states (see legend) measured via Raman spectroscopy. The matching solid lines result from the corresponding numerical simulation (see text).}
    \label{fig:evolution_interferogram}
\end{figure}

In Figure~\ref{fig:evolution_interferogram}, we show the interference fringes obtained for each individual momentum state in first ($n=1$) and third ($n=3$) order Bragg interferometers of $T=\qty{1}{\milli\second}$ (symbols) without OCT.
The $n=1$ case (panel a) manifests a clean sinusoidal pattern of two-state interference, dominated with more than 90\% amplitude by the principal frequency given by the target states $0\hbar k$ and $2\hbar k$. On the contrary, the observed interferometric fringes of target states $0\hbar k$ and $6\hbar k$ in case of $n=3$ interferometer (panel b) contain multiple harmonics. Only about 60\% resides in the principal frequency while the signal is strongly contaminated by the $n=1$ ($\approx25\%$) and $n=2$ ($\approx10\%$) parasitic interferometric loops. Here, the excellent momentum-state selectivity of the detection with Raman spectroscopy method underlines the breakdown of two-wave interference concept, which would complicate the single-shot extraction of interferometric phase from the observed probability.

The data is supported with simulation (solid lines) obtained by averaging the interferometer response over a Gaussian vertical velocity distribution of width $\sigma_v=0.3\,v_r$ and coupling inhomogeneities determined from the single pulse measurements (Table~\ref{tab:optimized_value}). 
For the first order Bragg interferometer, the Gaussian pulses are at resonance, with a peak Rabi frequency of $\qty{21.9}{\kilo\hertz}$ (Table~\ref{tab:optimized_value}). For the third order Bragg interferometer, the pulses had to be detuned by $- \qty{4.5}{\kilo\hertz}$ to reproduce the final population difference between the external states $-2\hbar k$ and $8 \hbar k$. The peak Rabi frequency is $\qty{72.6}{\kilo\hertz}$, which is consistent with the values obtained from the single pulse measurements (Table~\ref{tab:optimized_value}). 
In both cases shown in Figure~\ref{fig:evolution_interferogram}, the simulated fringes are "scanned" using a phase increment imprinted during the final beamsplitter pulse. The empiric conversion to a frequency ramp is achieved by linear fit of the phase shift to the frequency ramp to account for the corrections due to the finite pulse duration and its Gaussian shape. This contribution modifies the oscillation frequencies within a few percent, giving a fine match of horizontal scaling.
Our method, in principle, can be applied to analyze any multi-port atom interferometer, provided a sufficient spectroscopic resolution. 

\section{Conclusion}
In this work, we have presented a method for analyzing atom optics dynamics in the quasi-Bragg regime using Raman spectroscopy. This technique addresses a typical resolution limit for standard time-of-flight detection scheme, where the spatial separation of  adjacent Bragg states remains insufficient with respect to atomic cloud dispersion. By using velocity-selective Raman pulses, we successfully reconstructed the time evolution of atomic populations in momentum space during the interrogation pulses. 

For comparison with experimental data, we performed numerical simulation based on resolving the Schrödinger equation for atom-light interaction in quasi-Bragg regime. Our analysis demonstrated the importance of coupling inhomogeneities coming from the convolution of the intensity variation across the laser beam profile with the spatial distribution of the atomic cloud.

This tool offers a robust diagnostic of the Bragg dynamics, and allows us to track the complex evolution of the quantum state driven on multi-photon transitions, enhanced with optimal quantum control protocols in the presence of inhomogeneities. To achieve good agreement between theory and experiment, it was necessary to adjust the peak Rabi frequency and effective cloud size as fitting parameters, revealing that the atomic cloud experiences a broader than expected distribution of Rabi coupling.

In addition, we extended this characterization to a full Mach-Zehnder interferometer sequence. The spectral resolution of our method allowed for observing interference fringes specific to each Bragg state and quantifying the contribution of unwanted momentum states into reduced contrast of the main interferometric loop. 
\begin{acknowledgments}
We thank V. Cambier and F. Correia for their contribution at the earlier stage of this work, J. Pinto, L. Volodimer and M. Lours for general support in electronics, and, in particular, for design and implementation of the high-voltage amplifiers that play a key role in OCT measurements, and Q. Beaufils for careful reading of the manuscript. All authors acknowledge support from the Agence Nationale de la Recherche under Contract No. ANR-19-CE47-0003 GRADUS, and from a government grant managed by the Agence Nationale de la Recherche under the Plan France 2030 with the reference “ANR-22-PETQ-0005” (project QAFCA).
\end{acknowledgments}

\appendix

\section{Evolution of the Bragg states for various Bragg pulses}
\begin{figure}[htbp!]
    \centering
        \centering
        \includegraphics[width=\linewidth]{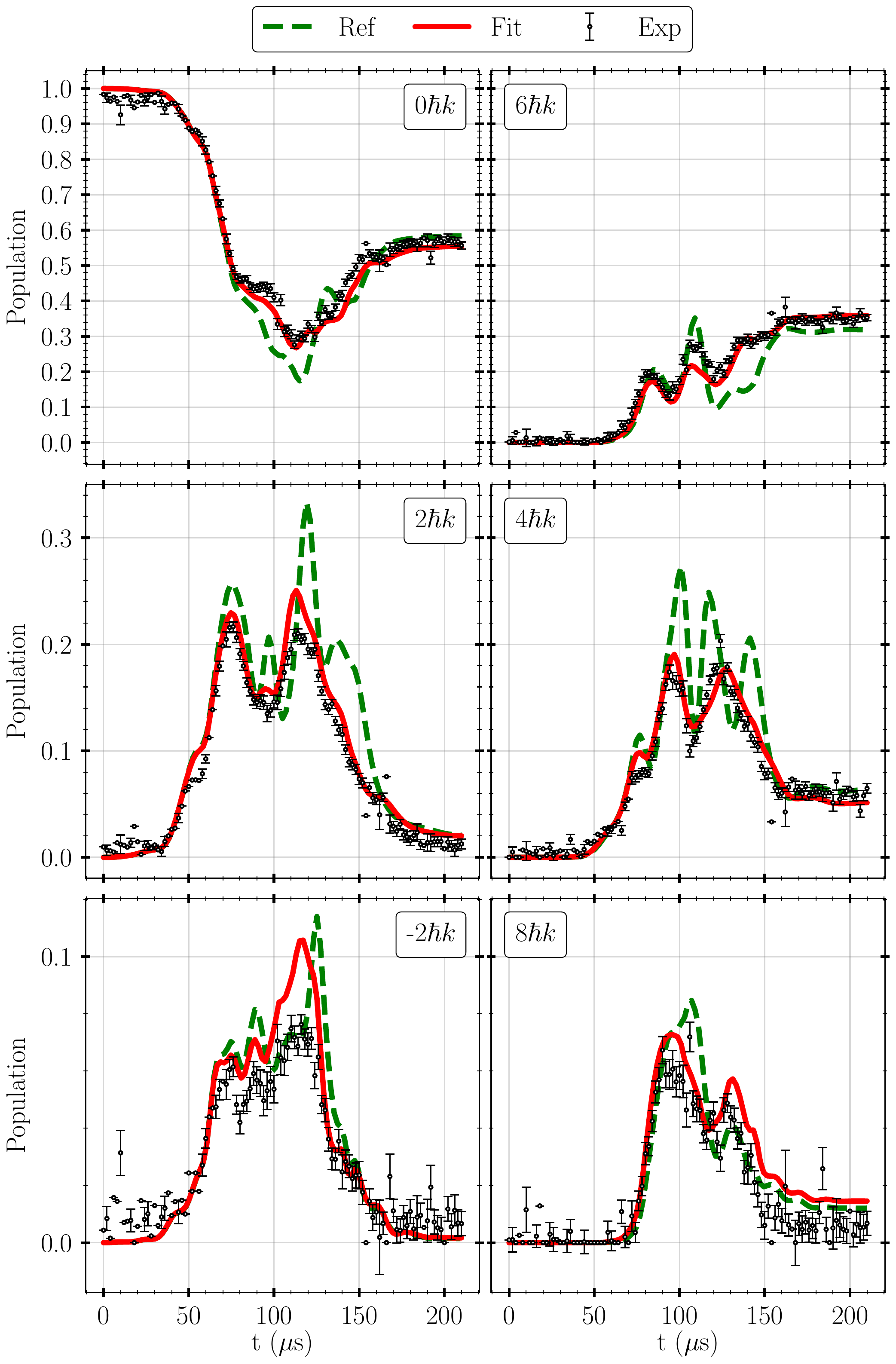}
        \caption{Evolution of the momentum states in the $6\hbar k$ Bragg beamsplitter pulse optimized with OCT, recorded using stroboscopic Raman spectroscopy (symbols). Dashed green (solid red) line is numerical simulation for the reference (fitted) parameter set, see main text.}
        \label{fig:evolution_spectro_pi_over_2_pulse_6hk_OCT}
\end{figure}
For a more complete demonstration of stroboscopic Raman spectroscopy method, we provide here further data, corresponding to the cases not shown in the main text for brevity, while being listed in Table~\ref{tab:optimized_value}. 
Figure~\ref{fig:evolution_spectro_pi_over_2_pulse_6hk_OCT} demonstrates the evolution of atomic state during $6\hbar k$ Bragg beamsplitter pulse enhanced with OCT protocol, supporting all conclusions outlined for the case of an optimized Bragg mirror pulse.

\begin{figure}[htbp!]
    \centering
        \centering
        \includegraphics[width=\linewidth]{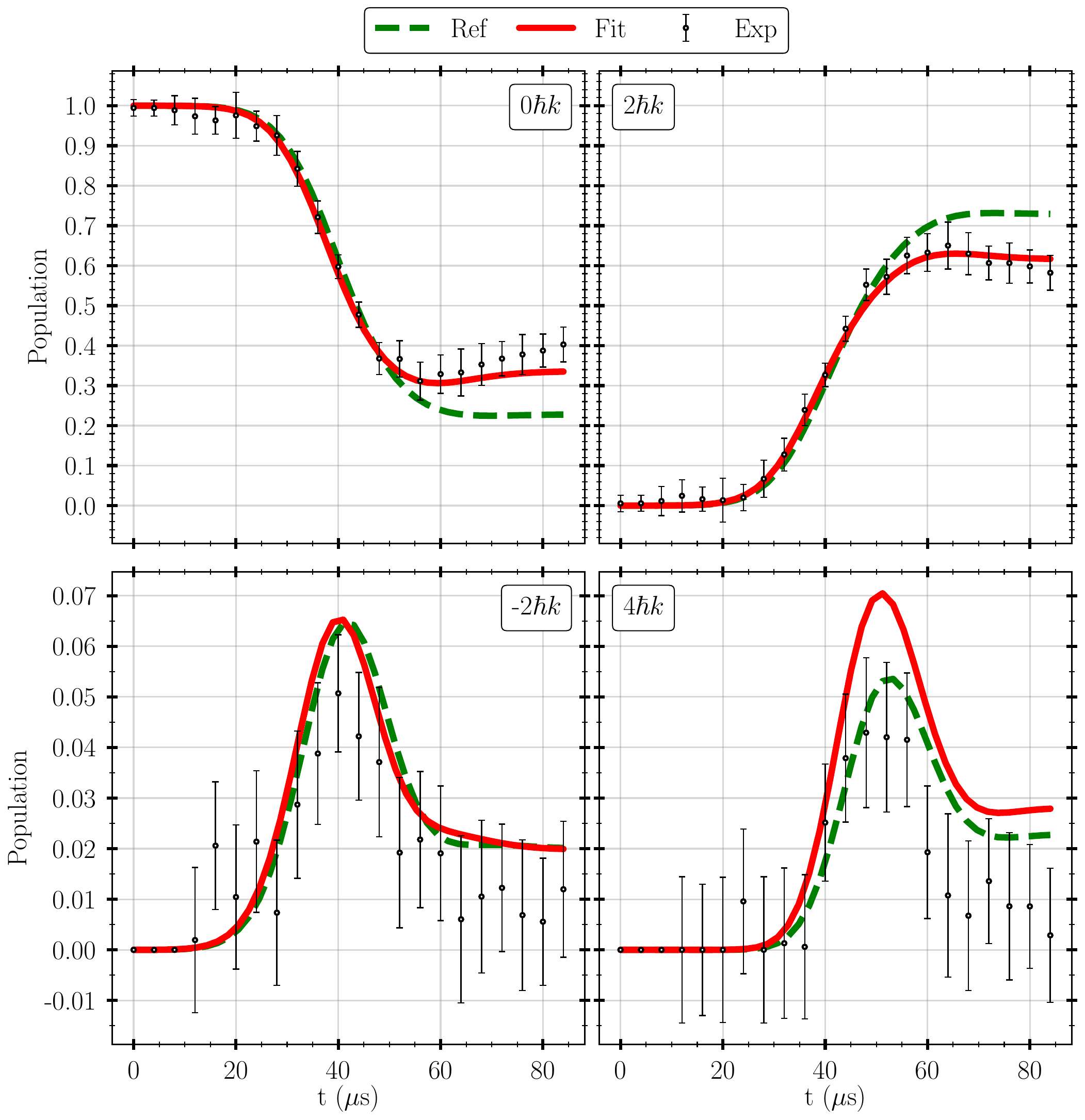}
        \caption{Evolution of momentum states in a conventional $n=1$ Bragg mirror pulse, recorded using stroboscopic Raman spectroscopy (symbols). Dashed green (solid red) line is numerical simulation for the reference (fitted) parameter set, see main text.}
        \label{fig:evolution_spectro_pi_pulse_2hk_no_OCT}
\end{figure}

In addition, we have applied the developed detection and simulation methods to conventional $2\hbar k$ Bragg pulses. Figure~\ref{fig:evolution_spectro_pi_pulse_2hk_no_OCT} shows corresponding recorded in-pulse evolution. Here, the coupling of unwanted states typically remains below 5\% - level that is nonetheless well resolved experimentally and matched with simulation. 

\section{Details on numerical simulation of the quasi-Bragg regime}

The Hamiltonian of an atom with a mass $M$ subject to the gravitational acceleration and a periodic potential created by counter-propagating laser beams of frequencies $\omega_1, \omega_2$ and wavevectors $k_1 \approx -k_2 \approx k$ along the $z$-direction can be written (in the free falling frame) as:
\begin{equation}
    \tilde{H} = \frac{\hat{p}^2}{2M} - \tilde{v}(t)\hat{p} - \hbar \Omega(t) \left( e^{2ik\hat{z}} + e^{-2ik\hat{z}} \right)
    \label{eq:H_eff}
\end{equation}
where $\tilde{v}(t)$ is the velocity of the lattice with respect to the free-falling atom:
\begin{equation}
    \tilde{v}(t) = \left(\frac{1}{2k}\frac{d\phi(t)}{dt}+gt\right)
    \label{eq:v_tilde}
\end{equation}
The periodicity of the lattice potential allows us to expand the wave function in the quasi-momentum basis: $\{\ket{p_0+2n\hbar k}\} \equiv \{\ket{2l\hbar k}\}$:
\begin{equation}
    \ket{\Psi(t)} = \sum_{n} c_l(t) \ket{2l\hbar k}
\end{equation}
where $l\in\mathbb{Z}, -2\leq l\leq 5$ is the Bragg state index that would formally coincide with diffraction order $n$ used in the main text as initial state is $l=0$.\\  
\\
\textbf{Simulation routine}\\
The simulation integrates an 8‑state system for each discrete class of velocity and (x,y) position using an internal RK4 solver (Numba JIT). For each point ($v_z,x,y$) the time evolution over the pulse window $t_0 \pm 3 \sigma_t$ pulse is computed, resulting in complex amplitudes $c_l(t)$ for the eight momentum states. The ($v_z,x,y$) points are then weighted by Gaussian velocity and spatial distributions and averaged to compute time‑dependent state probabilities and phases.
Numba routine performs the linear interpolation of an external phase profile, the evaluation of a time‑dependent reduced complex coupling including the Gaussian envelope and OCT phase $\gamma(t)\equiv\frac{\Omega_R(t)}{\Omega_0}\exp{(i\varphi_{\textrm{OCT}}(t))}$ with $\Omega_0/2\pi=\qty{14.25}{\kilo\hertz}$, and RK4 integration with adjustable internal sub-steps.\\
\\
\textbf{Discretization grids, weights and constraints}\\ 
We derive $21$ velocity classes $v_i$ from the range [$-1.5\,v_r$;$1.5\,v_r$] with step size of $0.15\,v_r$ and 23-per-axis discrete $(x_i,y_i)$ classes from the range [$-2\,\sigma_r$;$2\,\sigma_r$] with step size of $2\sigma_r/11$. Velocity weighting takes into account the velocity selection induced by a Raman square pulse of duration $\tau=70~\mathrm{\mu s}$; spatial weights are Gaussian functions of RMS width $\sigma_r$. The interrogating Gaussian beam of waist $w_0=\qty{4}{\mm}$ is considered to be centered on the atomic cloud.\\
The routine computes weighted averages of $|\psi|^2$ and phases over all points to produce the simulated dynamic.\\
\\
\textbf{Fitting procedure}\\
The fitted parameters are initial RMS cloud size $\sigma_r$ and adimensional coupling amplitude $\gamma_0$. The pipeline consists of: warm‑up JIT run → global optimization with differential evolution on a coarse simulation grid → local refinement with least‑squares on a finer grid.\\

At each iteration, the residuals $p_{\textrm{sim}}-p_{\textrm{exp}}$ are concatenated across all orders, while the orders with negligible signal ($-4\hbar k$ and $10\hbar k$) are eventually ignored setting the cutting threshold.
The fit residuals are not normalized by the uncertainties, in order to ensure equivalent treatment of all Bragg states and an absence of bias related to the possible accidental fluctuation in the error bars generated during the fits of spectral features (as those in Figure\ref{fig:Figure2} a and b).\\

We have observed that various pairs ($\Omega_R,\sigma_r$) can lead to the similar optimization result, suggesting a coupling between the two parameters. As these nearly-optimal pairs of parameters exceed the typical error bars from a single fit, we apply another common method for more robust error-bars estimation. Using the recorded experimental data, we generate for all states an array of $300$ simulated curves that include randomly sampled points from the normal distributions given by the mean value and error bar of each data point. We then fit the same model (as for single-fit protocol) with identical initialization conditions to this simulated data and obtain $300$ pairs of ($\Omega_R,\sigma_r$), from which we extract estimated mean and standard deviation for the two parameters.  

A possibility to constrain the coupling between $\Omega_R$ and $\sigma_r$ is to further increase the number of necessary parameters by including, for example, the RMS of vertical velocity distribution. Such extension may, however, demand significantly longer simulation times, which goes beyond the scope of present study.     


\clearpage
\bibliography{Joel_Raman}

\end{document}